\documentclass[preprint,prc,aps,floatfix,showpacs,showkeys]{revtex4}
\usepackage{epsfig}
\usepackage {graphicx}
\begin{document}
\title{Inclusive Cross-Sections of (p, xp) and (p, x$\alpha$)
Reactions on $^{56}$Fe Nucleus at E$_{p}$=29.9 MeV}
\author{A. Duisebayev and K. M. Ismailov}
\affiliation{Institute of Nuclear Physics, National Nuclear Center,
Republic of Kazakhstan}
\author{I. Boztosun}
\email{boztosun@erciyes.edu.tr} \affiliation{Department of Physics,
Erciyes University, Kayseri, Turkey}
\date{\today}
\begin{abstract}

In this paper, we present new experimental data measured at
E$_{p}$=29.9 MeV for the inclusive reactions (p,xp) and
(p,x$\alpha$) on nucleus $^{56}$Fe. The adequacy of the theoretical
models in explaining the measured experimental data is investigated
and the contributions of multi-step direct and multi-step compound
processes in the formation of the cross-sections are determined. It
should also be underlined that the traditional frameworks are valid
for the description of the new experimental data. A comparison with
the previous measurements for the (p,xp) and (p,x$\alpha$) on
$^{54}$Fe nucleus reveals that these data are in agreement with our
measurements. The only exception is within the energy region of
E$_{p}$=15 and 25 MeV for both reactions, where the cross-section
for the $^{56}$Fe nucleus is smaller than the cross-section for the
$^{54}$Fe nucleus.
%
%This observation shows that the essential isotope dependence of the
%reaction cross-sections is not observed.
\end{abstract}
 \maketitle

\section{INTRODUCTION}
The pre-equilibrium decay mechanism in nuclear reactions reflects
the dynamics of the formation of the excited system and its
evolution to the equilibrium state. Working out this mechanism
remains as an actual problem of the nuclear reaction theory. The
problem is largely connected with obtaining new experimental data on
double-differential cross-sections in (p,xp), (p,xd) {\it etc.}
Thus, it is anticipated that the availability of high-quality
experimental data on reactions with different proton energies
\cite{ref0}, using charged particles for the double-differential
cross-sections, could address this problem.

The reactions induced by protons within the energy range of 10-2000
MeV play a crucial role in applied research. Development of
electro-nuclear installations (Accelerator Driven System) for
nuclear transmutation, which arises from long-lived radioactive
waste of nuclear industry and energy production \cite{ref1}, is one
important example.

The physical scenario of such a system requires the presence of
experimental data on key parameters of nucleon interaction,
cross-sections of interaction, energy spectra and angle
distributions of secondary particles ($^{1,2,3}$H, $^{3,4}$He
\emph{etc.}). These particles can be agents initiating a reaction
with neutrons emission. The testing and perfection of the
theoretical methods and codes in describing the experimental
measurements become also very important.

In this study, the subject of the research is the $^{56}$Fe nucleus,
which is one of the basic constructional materials of hybrid
nuclear-energy installation. Early experimental studies on the
targets of $^{54,56}$Fe nuclei \cite{77,79,102,106,111,112,113} have
focused on the emission of protons, deuterons, and alpha particles.
The double differential cross section measurements, angle-integrated
spectra and energy-binned angular distributions obtained from these
experimental studies have been compared with the predictions of the
pre-equilibrium reaction theory (see a recent review by Koning and
Duijvestijn \cite{114} for a detailed discussion). Initial
experimental data on reactions (p,xp), (p,x$\alpha$) on nuclei of
isotope $^{54}$Fe at E$_{p}$=29.0 and 39.0 MeV have been measured in
the work of reference \cite{106} and the energy spectra of secondary
particles have been analyzed within the intra-nuclear cascade and
evaporation models. An acceptable description of the experimental
data have been achieved only for a spectrum higher than 20.0 MeV.

The Japanese group \cite{ref3} has investigated the cross-sections
of the reaction (p,xp) on targets $^{54,56}$Fe nuclei with a
thickness of 500 mg/cm$^{2}$ at the energy 26.0 MeV. The
experimental data have been analyzed within the framework of FKK
theory for pre-equilibrium processes, using the code FKK-GNASH and
within the framework of the Hauser-Feshbach model for compound
processes.

It is clear that the protons in the energy region of 30 MeV have not
been studied in detail. Extending the experiment in this direction
allows us to observe the mechanisms of the reaction and the level of
energy-dependence in detail and to use these observations for
adequate analysis within the framework of the pre-equilibrium
reaction theory.

Therefore, in our experiment we consider the (p,xp)$^{56}$Fe and
(p,x$\alpha$)$^{56}$Fe reactions at E$_{p}$=29.9 MeV within the
angle range of 30-135$^{\circ}$. In the following section, we
present our experimental method, the details of the measurement and
the experimental results. Section \ref{analysis} is devoted to the
theoretical analysis of the measured experimental data by the
exciton model and quantum mechanical representations. Finally,
Section \ref{conc} gives our summary and conclusion.

\section{EXPERIMENT AND RESULTS}

The experimental cross-sections measurement of reactions (p,xp) and
(p,x$\alpha$) have been carried out on a beam of the accelerated
protons at an energy of 29.9 MeV on the isochronous cyclotron,
U-150M, at the Institute of Nuclear Physics, NNC Republic of
Kazakhstan, by using a self-supporting target $^{56}$Fe. The
properties of the target nucleus are given in table \ref{tab1}. The
measurements have been executed within the angle range of
30-135$^{\circ}$ at intervals of 15$^{\circ}$ in the laboratory
system.

\begin{figure}[htbp]
\centerline{\includegraphics{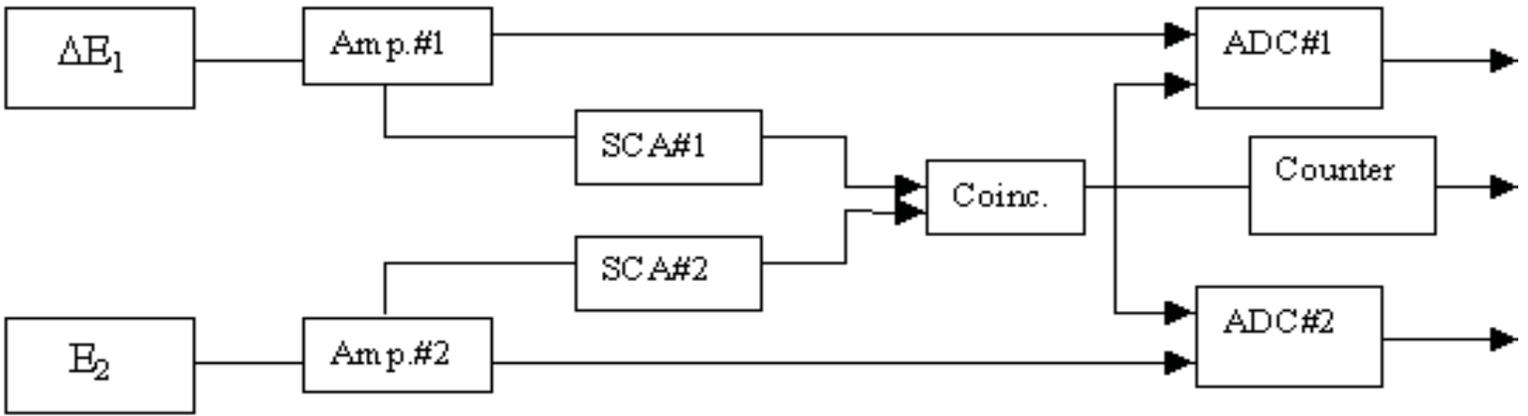}} \caption{The block-scheme of
the registration system: Amp.{\#}1,2- Spectroscopic Amplifiers;
SCA{\#}1,2- Single Channel Analyzers; Coinc. - Scheme of
Coincidences; Counter-Counter Scheme; ADC{\#}1,2- Analog - Digital
Converters.} \label{fig1}
\end{figure}

The registration and identification of the reaction products have
been carried out by a system of multi-programming analysis, based on
the use of the $\Delta$E-E-method, ORTEC and PC-spectrometric lines.
The block-scheme of the registration system is presented in Figure
\ref{fig1}. The detector telescope had a silicon surface-barrier
detector $\Delta$E with a thickness of 30 microns and E with a
thickness of 2000 microns for the reaction $^{56}$Fe(p,x$\alpha$).
Whereas for reaction $^{56}$Fe(p,xp), the thickness of the silicon
surface-barrier detector $\Delta $E was 500 microns and the
thickness of the stop detector of total absorption-crystal CsI(Tl)
was 25 mm. The solid angles of telescopes equalled to 2.72 10$^{-5}$
sr and 2.59 10$^{-5}$ sr respectively. The energy calibration of
spectrometers has been carried out by the kinematics of residual
nuclei levels in reaction $^{12}$C(p,x) and protons of recoil. The
total energy resolution of the system basically equalled to 400 keV,
and was determined by the energy resolution of the accelerated
protons beam. One should note that the real line spectra might be
distorted by the impurity of the light elements in the target
nucleus, accidental coincidences and background. Therefore, at each
angle, we measured the spectra both with and without target as well
as with the spectra of the light elements such as $^{12}$C and
$^{16}$O.

Thus, the systematic uncertainties were conditioned by the
uncertainties in determining the target thickness ($\sim$7$\%$), the
calibration of the current integrator ($\sim$1$\%$), and the solid
angle of the spectrometer ($\sim$1.3$\%$). The energy of accelerated
particles were measured accurately with 1.2$\%$. The uncertainties
of the registration angle were less than 0.5$\%$. The whole
systematic error was less than 10$\%$. The statistical uncertainties
at a long exposition time of the double-differential cross-sections
were less than 10$\%$ for protons and less than 20$\%$ for $\alpha$
particles in the high-energy region of the spectra.

\begin{figure}
\centerline{\includegraphics{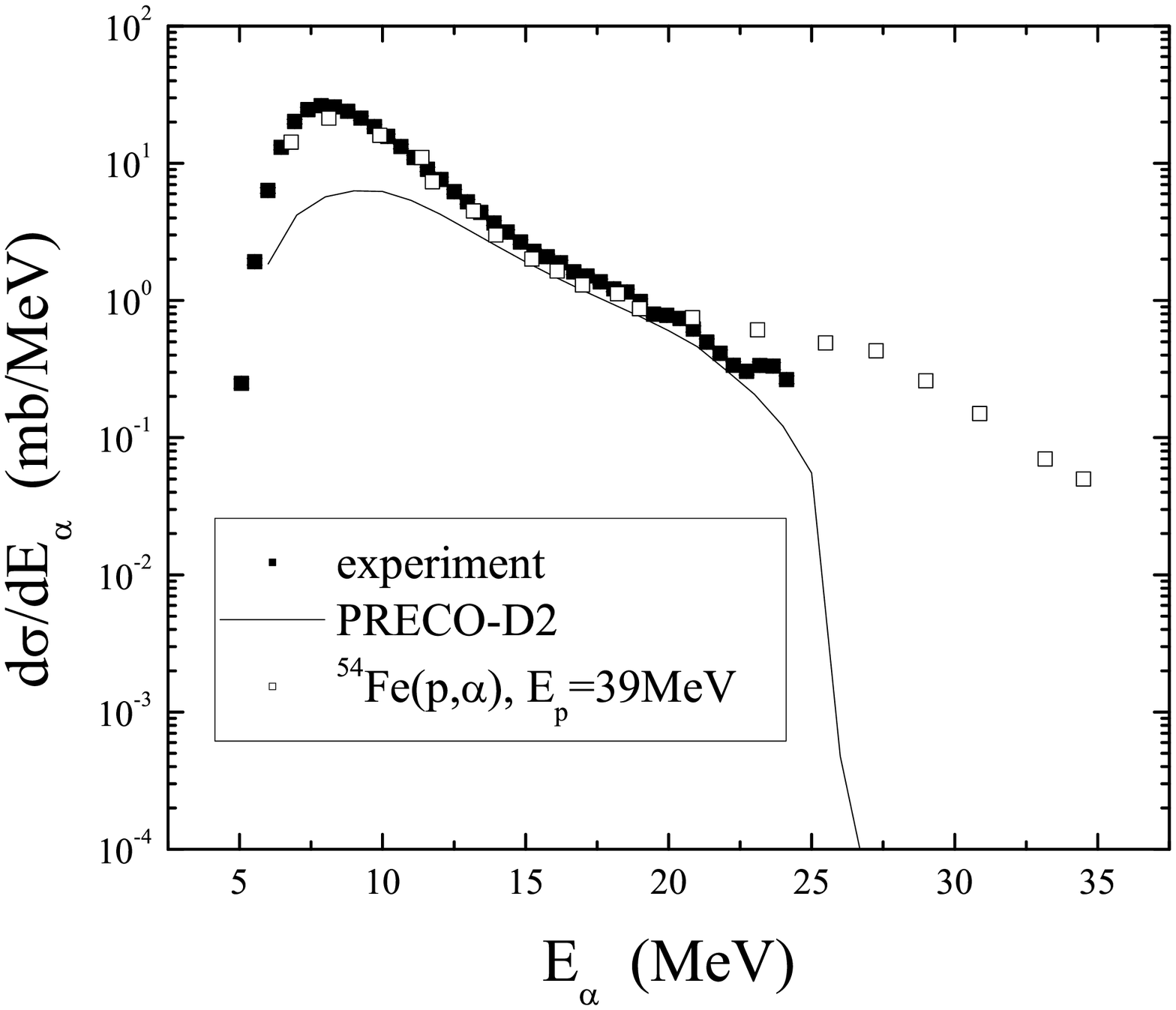}}\caption{The integral
cross-section of the $^{56}$Fe(p,x$\alpha )$ reaction at
E$_{p}$=29.9 MeV (filled squares). $^{54}$Fe(p,$\alpha )$ reaction
at E$_{p}$=39.0 MeV \cite{106} is also shown in comparison for the
isotope dependence of the reactions (empty squares).} \label{fig2}
\end{figure}

\begin{figure}
\centerline{\includegraphics{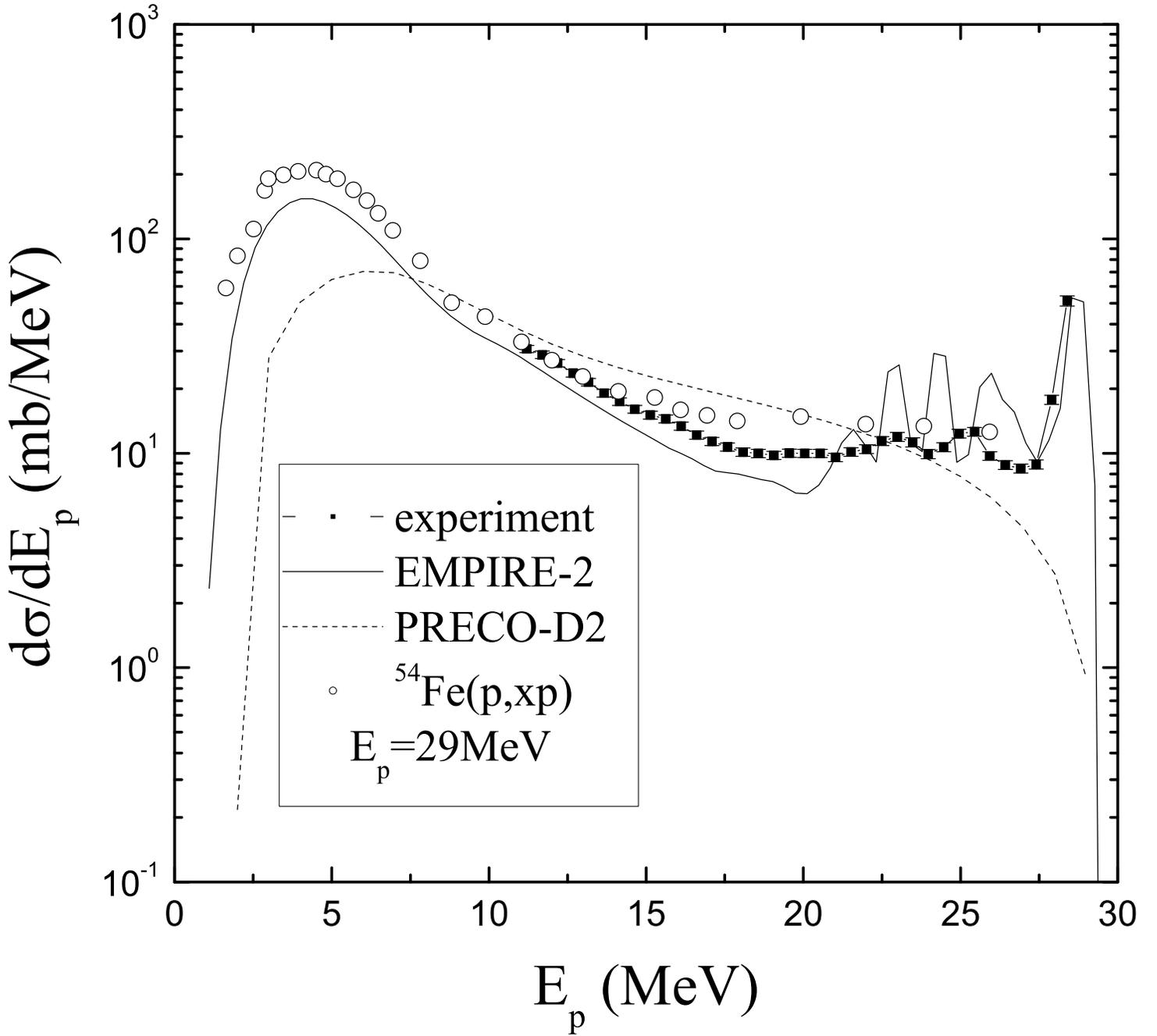}} \caption{The integral
cross-section of the $^{56}$Fe(p,xp) reaction at E$_{p}$=29.9 MeV
(filled circles). $^{54}$Fe(p,xp) reaction at E$_{p}$=29.0 MeV
\cite{106} is also shown in comparison for the isotope dependence of
the reactions (empty circles).} \label{fig3}
\end{figure}

The results of the measurements are shown in Figure \ref{fig2} for
the $^{56}$Fe(p,x$\alpha )$ reaction and  in Figure \ref{fig3} for
the $^{56}$Fe(p,xp) reaction together with analogical data on an
isotope of the iron nucleus $^{54}$Fe at 29.0 and 39.0 MeV,
presented in the work of reference \cite{106}. In accordance with
the measurement of reference \cite{106}, cross-sections of reactions
$^{56}$Fe(p,x$\alpha$) (Figure \ref{fig2}) and $^{56}$Fe(p,xp)
(Figure \ref{fig3}) practically coincide with the cross-sections of
the appropriate reactions on $^{54}$Fe. The exception is only in the
energy region of E$_{p}$=15-25 MeV for both reactions, where the
cross-section for the $^{56}$Fe is less than the cross-section for
the $^{54}$Fe nucleus. We have obtained the experimental partial
cross section by integrating the integral spectra (d$\sigma$/dE) on
energy. The experimental partial cross-sections of reactions
$^{56}$Fe(p,x$\alpha )$, (p,xp) are given in table \ref{tab2}.

\section{Analysis of the Results}
\label{analysis}

\begin{figure}
\centerline{\includegraphics{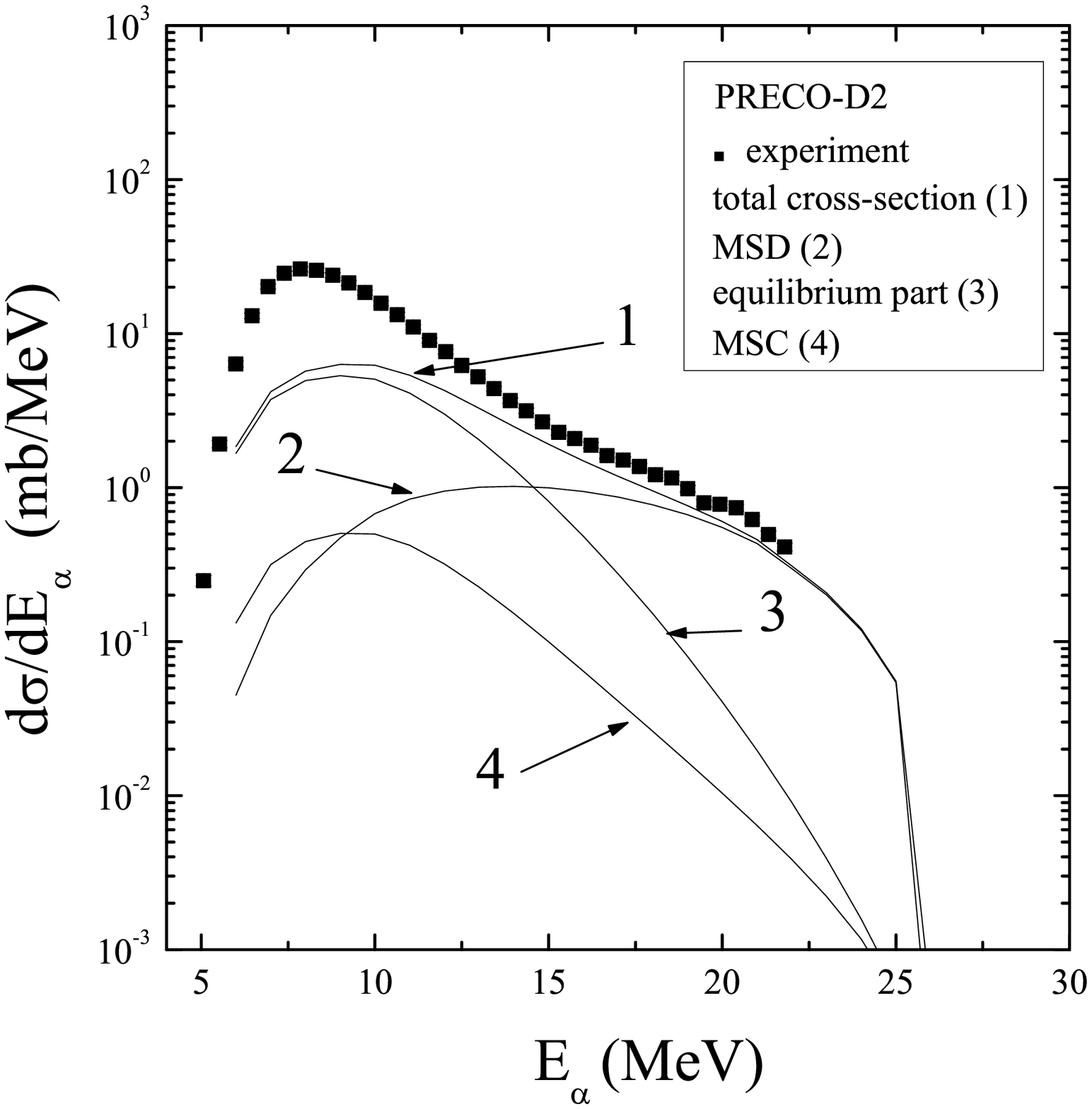}}\caption{The contribution MSD
and MSC mechanisms to the formation of the integral spectra of
reactions $^{56}$Fe(p,x$\alpha )$ at E$_{p}$=29.9 MeV obtained by
using the code PRECO-D2.} \label{fig4}
\end{figure}

\begin{figure}
\centerline{\includegraphics{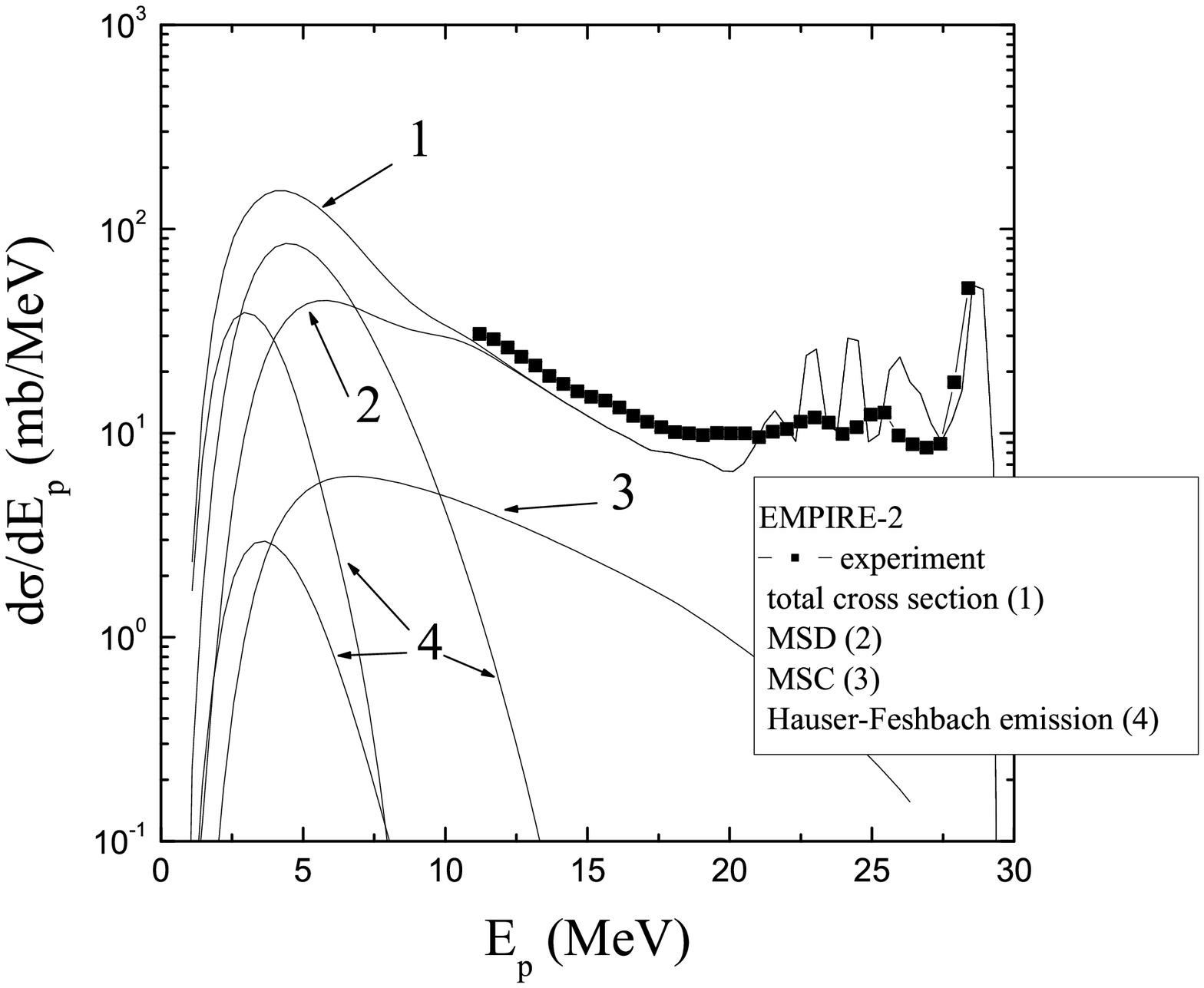}} \caption{The contribution
MSD and MSC mechanisms to the formation of the integral spectra of
reactions $^{56}$Fe(p,xp) at E$_{p}$=29.9 MeV obtained by using the
code EMPIRE-II.} \label{fig5}
\end{figure}

Many different theoretical approaches have been used to describe the
equilibrium reactions data over a wide range of incident energies
(see references \cite{114,ref3,Feshbach,Hodgson,Bonetti1,Bonetti2}
for a detailed discussion). In this paper, the analysis of the
experimental results has been conducted in the Griffin exciton model
\cite{ref4} of the pre-equilibrium decay of nuclei. The program
PRECO-D2 \cite{ref5}, which describes the emission of particles with
mass numbers from 1 up to 4, has been used in our theoretical
calculations. The Griffin exciton model is a statistical model where
the excited levels of the intermediate system are described in terms
of the single-particle shell model, \emph{i.e.} characterized by the
number of the excited particles (above the Fermi level) and holes
(below the Fermi level). It is assumed that the evolution of the
system occurs through a sequence turning into complicated
configurations and the emission of particles is possible on each
phase of this evolution. The conditions of the intermediate system
are divided into two classes; bound and unbound. This allows the
calculation of the integrated cross-sections  on angle for the
statistical multi-step direct (MSD) and multi-step compound (MSC)
processes \cite{ref6} in the exciton model. The calculated
contributions of the MSD and MSC processes in the formation of the
total cross-section of reactions $^{56}$Fe(p,xp), (p,x$\alpha )$ are
shown in figures \ref{fig4} and \ref{fig5}. The contribution of
additional MSD components that are not taken into account by the
Griffin model have been determined semi-empirically by taking into
account the direct nucleon transfer reaction and knock-out direct
processes, including cluster freedom degrees. The evaporation from
the equilibrium state of the nucleus has been included in the total
cross-section. The configuration (1p0h) has been accepted as the
initial particle-hole configuration in all calculations.

The density of the particle-hole states is given by

\begin{equation}
\omega (p,h,E) = \frac{g(gE - A_{ph} )^{n - 1}}{p!h!(n - 1)!},
\end{equation}
where

\begin{equation}
A_{ph} = \frac{(p^2 + h^2) + (p - h) - 2h}{4}
\end{equation}

The single-particle density of levels has been accepted as g=A/13.

The optical potential parameters of Huizenga \cite{ref7} for
$\alpha$-particles and of F. D.Becchetti and G. W.Greenlees
\cite{ref8} for protons have been used.

The comparison of the experimental results and theoretically
calculated spectra is shown in Figures \ref{fig4} and \ref{fig5}. It
can be seen from these figures that the basic contribution to the
hard part of the total cross-section is caused by the MSD mechanism.
It is also observed that the evaporated part of the cross-section is
underestimated within the framework of the exciton model used. This
may be due to the fact that the preferred approach gives only the
pre-equilibrium part of the MSC process without taking into account
the emission from complex equilibrium configuration of the compound
system. Therefore, the analysis of the experimental cross-sections
of the $^{56}$Fe(p,xp) reaction is carried out within the
Hauser-Feshbach theory by considering the multi-particle emission of
both single-charged (protons, deuterons) and two-charged fragments
($\alpha$-particles) by using the program EMPIRE-II \cite{ref9}. In
this code, the contributions of statistical direct and compound
processes are described by the optical model (SCAT 2 \cite{ref10}),
multi-step direct (ORION+TRISTAN \cite{ref11,ref12} and multi-step
compound (NVWY \cite{ref13}) models. The parameter of the level
density has been defined by the Gilbert-Cameron parameterizations
\cite{ref14}.

The results of the calculations using the Hauser-Feshbach theory are
given in table \ref{tab3} and are shown in Figures \ref{fig4} and
\ref{fig5}. These results display that the contribution of the
multi-particle compound mechanism determines the emission of protons
from 2.5 MeV up to 10 MeV and that the contribution of multi-step
direct process ranges from 5 MeV up to the kinematical limit. The
form of integral spectra of reaction (p,xp) is determined by the
multi-step direct processes.

\section{SUMMARY AND CONCLUSIONS}
\label{conc}

We have measured the experimental data at E$_{p}$=29.9 MeV  within
the angle range of 30-135$^{\circ}$ for the inclusive reactions
(p,xp) and (p,x$\alpha$) on nucleus $^{56}$Fe, which has not been
investigated in detail so far. We have shown the extension of the
pre-equilibrium reactions to this energy region and have interpreted
the results of the experiments. The adequacy of the theoretical
models in explaining the measured experimental data is also
discussed. In our theoretical analysis, the contributions of
multi-step direct and compound processes in the formation of
cross-sections are determined and we assert that the traditional
frameworks are valid for the description of the experimental data.

%\newpage

\begin{table}[htbp]
\begin{center}
\begin{tabular}{|c|c|c|}\hline
 & Thickness, mg/cm$^{2}$ & Enrichment, {\%}  \\\hline
$^{56}$Fe & 2.7 & 95  \\\hline
\end{tabular}
\caption{Characteristics of the target nucleus.} \label{tab1}
\end{center}
\end{table}

\begin{table}[htbp]
\begin{tabular}{|c|c|c|}
%{|p{142pt}|p{142pt}|p{142pt}|}
\hline $^{56}$Fe& Energy range (MeV)&
$\sigma $(mb) \\
\hline (p,x$\alpha )$& 5-25&
132.2$\pm $1.2  \\
\hline (p,xp)& 11-27&
219.41$\pm $1.4 \\
\hline
\end{tabular}
\caption{The experimental partial cross-section of the reactions
(p,x$\alpha )$ and (p,xp).} \label{tab2}
\end{table}

\begin{table}[htbp]
\begin{tabular} %{lllllllll}
{|p{78pt}|p{56pt}|p{56pt}|p{56pt}|p{56pt}|p{85pt}|p{70pt}|} \hline &
Energy range (MeV)& Total cross- section (mb)& MSD cross- section
\par (mb)& MSC cross- section \par (mb)& Equilibrium cross-section
(mb) &
Hauser-Feshbach emission (mb) \\
\hline PRECO-D2  \par $^{56}$Fe (p,x$\alpha )$& 6-28& 46.8 & 11.3&
3.2& 32.24&
- \\
\hline EMPIRE-II \par $^{56}$Fe (p,xp)& 1-30& 1083.6& 559.2& 64.9&
-&
459.5 \\
\hline
\end{tabular}
\caption{The theoretical contributions of various mechanisms forming
the total cross-sections of reactions (p,x$\alpha )$ and (p,xp).}
\label{tab3}
\end{table}



\end{document}